\documentclass[3p,times,procedia]{elsarticle}
\usepackage{nupha_ecrc}

\volume{00}

\firstpage{1}

\journalname{Nuclear Physics A}

\runauth{}

\jid{nupha}

\jnltitlelogo{Nuclear Physics A}

\usepackage{amssymb,amsmath}
\usepackage[figuresright]{rotating}

\begin{document}

\begin{frontmatter}



%% Instructions from Editor: Please use the following \dochead only in the preprint version (e-print arXiv etc.); 
%% use empty \dochead{} when submitting to Nuclear Physics A!
\dochead{XXVIIth International Conference on Ultrarelativistic Nucleus-Nucleus Collisions\\ (Quark Matter 2018)}
%% e.g. \dochead{17th International Conference on Dynamical Processes in Excited States of Solids}

\title{Rapidity decorrelation from hydrodynamic fluctuations}


\author{Azumi Sakai}
\author{Koichi Murase}
\author{Tetsufumi Hirano}

\address{Department of Physics, Sophia University, Tokyo 102-8554, Japan}

\begin{abstract}

We discuss %the physics of 
rapidity decorrelation caused by hydrodynamic fluctuations
in high-energy nuclear collisions at the LHC energy.
We employ an integrated dynamical model which is a combination of the Monte Carlo version of Glauber model with
extension to longitudinal direction for initial conditions, 
full three-dimensional relativistic fluctuating hydrodynamics for the space-time evolution of created matter 
in the intermediate stage and a hadronic cascade model in the late stage.
We switch on and off the hydrodynamic fluctuations in the hydrodynamic stage 
to understand the effects of hydrodynamic fluctuations on factorisation ratios $r_{n}(\eta^a_p, \eta^b_p)$. 
To understand the rapidity gap dependence of the factorisation ratio comprehensively,
we analyse Legendre coefficients $A^k_2$ and $B^k_2$.
\end{abstract}


\end{frontmatter}

% \linenumbers

\section{Introduction}
\label{Introduction}

In high-energy nuclear collisions, 
the deconfined nuclear matter, the quark gluon plasma (QGP), is
created. Just after the collisions, the QGP fluid is formed, expands, cools down and turns into hadron gases. Since final hadronic observables reflect the whole evolution of the system, it is important to construct integrated dynamical models to describe each stage of collisions. 
At each stage, fluctuations plays an important role in final observables.
For example, initial fluctuations of the geometry cause anisotropic flow.
Thermal fluctuations during expansion of the QGP fluids would also 
affect hadronic observables.
In this study we focus on the thermal fluctuations in the hydrodynamic stage also known as hydrodynamic fluctuations and investigate their effects on correlation function of anisotropic flow coefficients in the rapidity direction.

\section{Model, settings and analysis}
\label{Model, settings and analysis}

We employ an integrated dynamical model~\cite{model,muraseD,Murase:2016rhl} %in
on an event-by-event basis. 
This model is a combination of three models corresponding to each stage
of high-energy nuclear collisions.
For the initial conditions, the Monte-Carlo version of the Glauber model,
extended to longitudinal direction based on the modified BGK model
to capture the wounded nucleon picture \cite{initial}, is employed.
Initial entropy density distributions are obtained by using this model \cite{model}.
In the intermediate stage, we describe the space-time evolution of the created matter
by relativistic fluctuating hydrodynamics. 
The dynamical equation for shear stress tensor $\pi^{\mu\nu}$ in the second-order fluctuating hydrodynamics is 
\begin{align}
\label{eq:shear_stress_tensor}
\tau_\pi{\Delta^{\mu\nu}}_{\alpha\beta}u^{\lambda}\partial_{\lambda}\pi^{\alpha\beta}+\pi^{\mu\nu}\left(1+\frac{4}{3}\tau_\pi\partial_{\lambda}u^\lambda \right) &= 2\eta{\Delta^{\mu\nu}}_{\alpha\beta}\partial^\alpha u^\beta+\xi^{\mu\nu}.
\end{align}
Here $\eta$ is shear viscosity, $u^{\mu}$ is four fluid velocity and 
${\Delta^{\mu\nu}}_{\alpha\beta}=\frac{1}{2}\left({\Delta^\mu}_\alpha{\Delta^\nu}_\beta+{\Delta^\mu}_\beta{\Delta^\nu}_\alpha \right)-\frac{1}{3}{\Delta^{\mu\nu}}\Delta_{\alpha\beta}$. Relaxation time $\tau_{\pi}$ is included so that the system obeys causality.
To include the fluctuation term $\xi^{\mu \nu}$, we consider fluctuation dissipation relations
\begin{gather}
\label{eq:FD}
\langle\xi^{\mu\nu}(x)\xi^{\alpha\beta}(x')\rangle = 4\eta T\Delta^{\mu\nu\alpha\beta}\delta_\lambda^{(4)}(x-x'),
\\
\delta_\lambda^{(4)}(x-x') = \frac{1}{\Delta t} \frac{1}{(4\pi \lambda^2)^\frac{3}{2}} \exp{\left[-\frac{(x-x')^2}{4\lambda^2}\right]}.
\end{gather}
Here $\delta_\lambda^{(4)}(x-x')$ is a smeared delta function
and $\lambda$ is a Gaussian width. For an equation of state, we employ \textit{s}95\textit{p}-v1.1~\cite{EOS} which is a combination of a lattice QCD in high temperature and a hadron resonance gas in low temperature.
Shear viscosity is $\eta/s=1/4\pi$ and the relaxation time is $\tau_\pi= 3/4\pi T$.
We switch description from hydrodynamics to kinetic theory through the Cooper Frye formula \cite{CooperFrye} with the switching temperature $T_{\mathrm{sw}}=155$ MeV\@.
%\@. means end of sentence period
The subsequent evolution of the hadronic gas is described by a hadron cascade model JAM~\cite{Nara:1999dz}.

To investigate the effect of hydrodynamic fluctuations, we focus on 
the factorisation ratio in the longitudinal direction and its Legendre coefficients.
The factorisation ratio is defined as 
\begin{align}
\label{eq:factorisation_ratio}
r_n(\eta^a_p,\eta^b_p) &= \frac{V_{n\Delta}(-\eta^a_p,\eta^b_p)}{V_{n\Delta}(\eta^a_p,\eta^b_p)},&  V_{n\Delta} &= \langle{\cos (n\Delta\phi)} \rangle.
\end{align}
Here $V_{n\Delta}$ is the Fourier coefficient of two particle correlation functions.
$\Delta \phi $ is the difference of azimuthal angle between two particles.
These two particles are taken from the two separated region. 
When the factorisation ratio is close to unity, the event plane angle is aligned 
in the longitudinal direction.
On the other hand, when the event plane angle depends on rapidity, 
the ratio becomes smaller than unity.

For a further study of rapidity decorrelation, we propose the Legendre coefficient for rapidity dependence of anisotropic flow.
For each hydrodynamic event, we perform multiple hadronic cascade simulations
and average all these events to obtain the magnitude of anisotropic flow coefficient $v_n$ and its event plane angle $\Psi_n$.
The Legendre coefficients are obtained on an event-by-event basis
by performing the Legendre expansion of $v_n$ and $\Psi_n$,
\begin{align}
v_n(\eta_p) &= \sum_{k=0}^\infty a_n^k P_k\left(\frac{\eta_p}{\eta_\mathrm{max}}\right),&  \Psi_n
(\eta_p) &= \sum_{k=0}^\infty b_n^k P_k\left(\frac{\eta_p}{\eta_\mathrm{max}}\right),
\end{align}
where $a_n^k$ are the Legendre coefficients of $v_n$ and $b_n^k$ are of $\Psi_n$ and the rapidity window is $|\eta_\mathrm{max}|=2.5$.
$P_k$ are the Legendre polynomials and the first three are

\begin{equation}
P_0\left(\frac{\eta_p}{\eta_\mathrm{max}}\right) = 1, \quad
P_1\left(\frac{\eta_p}{\eta_\mathrm{max}}\right) = \frac{\eta_p}{\eta_\mathrm{max}}, \quad
P_2\left(\frac{\eta_p}{\eta_\mathrm{max}}\right) = \frac{1}{2}\left[3\left(\frac{\eta_p}{\eta_\mathrm{max}}\right)^2-1\right].
\end{equation}

\section{Results}
\label{Results}
The number of hydrodynamic events is 4000 and the number of oversampling for hadronic cascade simulation is 100 for each hydrodynamic event.
In the following, we show the results for ideal hydrodynamics, viscous hydrodynamics and fluctuating hydrodynamics with $\lambda =$ 1.0 fm and $\lambda =$ 1.5 fm for comparison.
By using the phase space distributions from the simulations,
we analyse the factorisation ratio $r_2(\eta^a_p,\eta^b_p)$ and the Legendre coefficients to understand rapidity decorrelation caused by hydrodynamic fluctuations.

\begin{figure}[htbp]
\centering
\includegraphics[width=0.6\textwidth]{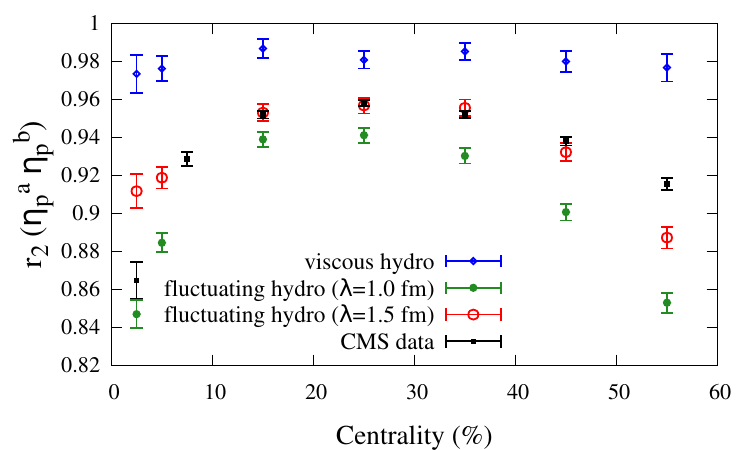}
\caption{Centrality dependence of factorisation ratio $r_2(\eta^a_p,\eta^b_p)$.
Rapidity regions are $2.0<\eta^a_p<2.5$ and $3.0<\eta^b_p<4.0$.
Results for viscous hydrodynamics (filled diamond), fluctuating hydrodynamics with $\lambda = 1.0$ fm (filled circle), fluctuating hydrodynamics with $\lambda = 1.5$ fm (open circle) and CMS data (filled square)~\cite{rn:CMSdata} are shown. 
}
\label{fig:one}
\end{figure}
Figure~\ref{fig:one} shows
results of centrality dependence of the factorisation ratio $r_2(\eta^a_p,\eta^b_p)$ 
compared with the CMS data \cite{rn:CMSdata}.
Results from fluctuating hydrodynamics are always smaller than viscous hydrodynamics.
In central collisions (0--10\% centrality),
fluctuating hydrodynamics with $\lambda = 1.0$ fm is close to experimental data.
On the other hand, in the mid-central collisions (10--40\% centrality) fluctuating hydrodynamics with $\lambda = 1.5$ fm is close to CMS data.
In comparison with the result from viscous hydrodynamics, 
hydrodynamic fluctuations tend to break factorisation.
If we choose the proper value for $\lambda$ at each centrality, we could reproduce factorisation ratios from the experimental data.

\begin{figure}[htbp]
\centering
\includegraphics[width=0.45\textwidth, bb=40 0 330 210]{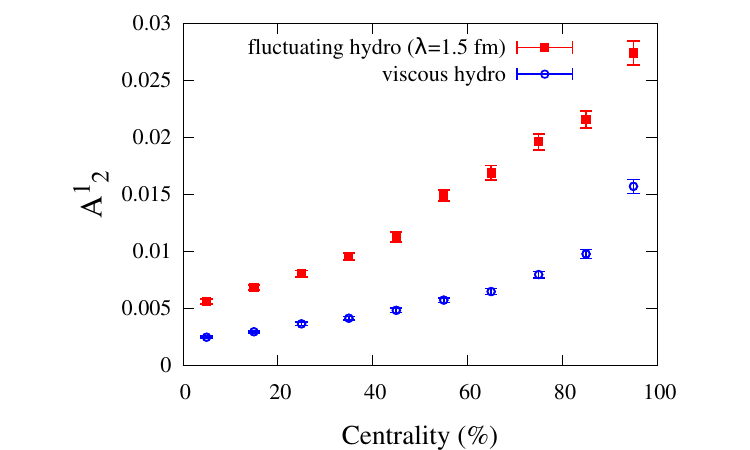}%
\includegraphics[width=0.45\textwidth, bb=40 0 330 210]{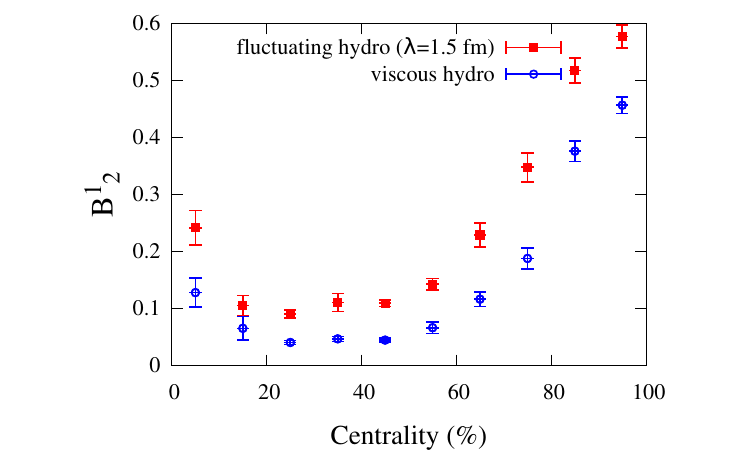}
\caption{Centrality dependence of the first-order Legendre coefficient  $A^1_2$
and $B^1_2$
for $v_2$ (left figure) and $\Psi_2$ (right figure), respectively. Results for viscous hydrodynamics (filled diamond) and fluctuating hydrodynamics with $\lambda = 1.5$ fm (filled square) are shown. 
}
\label{fig:two}
\end{figure}
To further understand the rapidity decorrelation more comprehensively, we calculate the Legendre coefficients for rapidity dependence of anisotropic flow. 
For each centrality, we calculate their root mean squares,
\begin{align}
A_n^k &= \sqrt{\frac{1}{N_\mathrm{ev}}\sum^{N_\mathrm{ev}}_{k=1}(a_n^k)^2},&
B_n^k &= \sqrt{\frac{1}{N_\mathrm{ev}}\sum^{N_\mathrm{ev}}_{k=1}(b_n^k)^2}.
\end{align}
Here $N_\mathrm{ev}$ is the the number of hydrodynamic events in that centrality.
Figure~\ref{fig:two} shows
results of the first-order Legendre coefficients for $v_2$ and $\Psi_2$ as a function of centrality.
Results in fluctuating hydrodynamics are always larger than the ones in viscous hydrodynamics. 
This means hydrodynamic fluctuations give
larger rapidity dependence of the magnitude of anisotropic flow coefficients and the event plane angle.

\begin{figure}[htbp]
\centering
\includegraphics[width=0.6\textwidth]{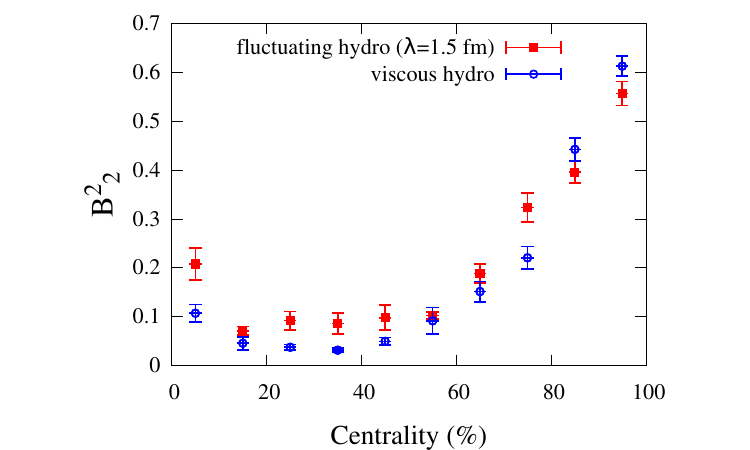}
\caption{Centrality dependence of second-order Legendre coefficient for event plane angle $\Psi_2$ ($B^2_2$). Results for viscous hydrodynamics (filled diamond) and fluctuating hydrodynamics with $\lambda = 1.5$ fm (filled square) are shown. 
}
\label{fig:three}
\end{figure}

We also analyse the second-order Legendre coefficients for the event plane angle
and results are shown in Fig.~\ref{fig:three}.
We call this $B^2_2$ ``second-order twist" of the event plane.
The second-order twist shows that the event plane angle decreases and increases again
along rapidity.
From the result of $B^2_2$, the second-order twist is finite and large, in particular, in central and peripheral collisions.
Also it tends to be larger in fluctuating hydrodynamics than in viscous hydrodynamics. 
The twist behavior of the event plane angle is often discussed through the first-order coefficient.
However, higher-order twists are also important in understanding event plane decorrelations.

\section{Conclusion and discussions}
\label{Conclusion and discussions}
We analysed effects of hydrodynamic fluctuations by performing numerical simulations of an integrated dynamical model based on full three dimensional fluctuating hydrodynamics.
 We discussed the importance of hydrodynamic fluctuations by comparing the factorisation ratios with experimental results.
The fluctuating hydrodynamic model tends to break factorisation, and the factorisation ratios become closer to experimental results.
Therefore hydrodynamic fluctuations play an crutial role in understanding the factorisation ratios $r_{n}(\eta^a_p, \eta^b_p)$.

We then proposed the Legendre coefficients $A^k_2$ and $B^k_2$ to understand the effects of hydrodynamic fluctuations. By analysing the Legendre coefficients using the fluctuating hydrodynamic model, we found hydrodynamic fluctuations tend to increase rapidity dependence. In this analysis, we found the second-order twist is large in central and peripheral collisions. Therefore higher-order twists are also important in understanding event plane decorrelations.

\section*{Acknowledgment}
This work was supported by JSPS KAKENHI Grant Numbers JP18J22227(A.S.) and JP17H02900(T.H.).



%% References


\bibliographystyle{elsarticle-num}
\bibliography{<your-bib-database>}



\end{document}